# Giant plasmonic bubbles nucleation under different ambient pressures

Binglin Zeng,[1,2,3,4] Yuliang Wang,[1,2,3,*] Mikhail E. Zaytsev,[2,4] Chenliang Xia,[1]
Harold J. W. Zandvliet,[4,†] and Detlef Lohse[2,5,‡]

[1]*School of Mechanical Engineering and Automation, Beihang University, 37 Xueyuan Rd, Haidian District, Beijing, China*
[2]*Physics of Fluids Group, Department of Applied Physics and J. M. Burgers Centre for Fluid Dynamics,
University of Twente, P.O. Box 217, 7500 AE Enschede, The Netherlands*
[3]*Beijing Advanced Innovation Center for Biomedical Engineering, Beihang University, 37 Xueyuan Rd, Haidian District, Beijing, China*
[4]*Physics of Interfaces and Nanomaterials, MESA+ Institute for Nanotechnology, University of Twente, 7500 AE Enschede, The Netherlands*
[5]*Max Planck Institute for Dynamics and Self-Organization, Am Fassberg 17, 37077 Göttingen, Germany*

Water-immersed gold nanoparticles irradiated by a laser can trigger the nucleation of plasmonic bubbles after a delay time of a few microseconds [Wang *et al.*, Proc. Natl. Acad. Sci. USA **122**, 9253 (2018)]. Here we systematically investigated the light-vapor conversion efficiency, $\eta$, of these plasmonic bubbles as a function of the ambient pressure. The efficiency of the formation of these initial-phase and mainly water-vapor containing bubbles, which is defined as the ratio of the energy that is required to form the vapor bubbles and the total energy dumped in the gold nanoparticles before nucleation of the bubble by the laser, can be as high as 25%. The amount of vaporized water first scales linearly with the total laser energy dumped in the gold nanoparticles before nucleation, but for larger energies the amount of vaporized water levels off. The efficiency $\eta$ decreases with increasing ambient pressure. The experimental observations can be quantitatively understood within a theoretical framework based on the thermal diffusion equation and the thermal dynamics of the phase transition.

## I. INTRODUCTION

Water-immersed noble-metal nanoparticles under irradiation of continuous-wave lasers can rapidly produce large amounts of heat when the plasmon resonance frequency of the nanoparticle matches with the laser frequency, resulting in the explosive boiling of water surrounding the nanoparticles. This explosive boiling results in the nucleation and growth of so-called plasmonic bubbles [1–6]. These plasmonic bubbles are of great importance in numerous plasmonic-enhanced applications, ranging from cancer therapeutics [7–10], catalytic reactions [11], micromanipulation of nano-objects [12–14], and solvothermal chemistry [1]. They also have been proposed for the conversion of solar energy [15–21]. In all these appli-cations, light-induced vapor formation plays a key role. How efficiently the light can be converted into vapor during this process remains, however, unclear. The light-vapor conversion efficiency is related to the growth dynamics of the plasmonic bubbles as well as the physicochemical properties of the sur-rounding liquid [22].

Previous studies on plasmonic bubble formation and growth dynamics have mainly focused on the milliseconds to seconds timescale [2,4,5,17]. Plasmonic bubbles formed on these timescales are hereafter referred to as ordinary plasmonic bubbles. In one of our previous studies we have shown that the growth of these ordinary plasmonic bubbles in water can be divided into two phases, a vaporization-dominated phase and a gas-diffusion dominated phase [23]. Plasmonic bubbles in the former phase have a smaller size. Water in the vicinity of the three-phase contact line is in direct contact with the laser spot. A relatively large fraction of the energy dumped in the nanoparticles is used to vaporize the surrounding water. As a result, these bubbles mainly contain vapor and exhibit a relatively high light-vapor conversion efficiency. In contrast, later the ordinary plasmonic bubbles contain both vapor and gas and are substantially larger. Therefore, the laser spots are then completely isolated from the water by the growing plas-monic bubbles [24]. Consequently, the heat at the laser spots cannot be directly transferred into the surrounding liquid. This significantly reduces the light-vapor conversion efficiency. As a result, the diffusion of dissolved gas expelled from the surrounding liquid dominates the growth of the plasmonic bubbles; consequently, they mainly contain gas and the light-vapor conversion efficiency of this phase is substantially lower than in the vapor-dominated phase.

We recently analyzed the very initial plasmonic bubble phase on a time scale of microseconds [6]. In this very initial phase a giant plasmonic bubble forms after a short delay time after switching on the laser, with a growth rate that is about three orders of magnitude larger than the ordinary plasmonic bubbles [6]. The lifetime of these initial phase plasmonic bubbles is, however, very short. Shortly after their formation they collapse due to the condensation of vapor [6].

The relatively large light-vapor conversion efficiency and the explosive growth rate of the giant initial plasmonic bubbles makes them very interesting for numerous applications. However, the underlying mechanism for their formation, as well as the light-vapor conversion process during bubble nucleation, are not quantitatively understood yet. Among the various physicochemical properties of the liquid such as the latent heat, obviously also the boiling point is very relevant for the nucleation and formation of plasmonic bubbles. However, it is very challenging to tune the boiling point of a liquid without changing the other physicochemical properties. Here we have varied the boiling point of water from 100 to 175°C by changing the ambient pressure from 1 to 9 bar. We have studied the nucleation and growth of the initial giant plasmonic bubbles under different ambient pressures and laser powers in order to obtain a thorough and solid understanding of the bubble nucleation as well as the light-vapor conversion processes.

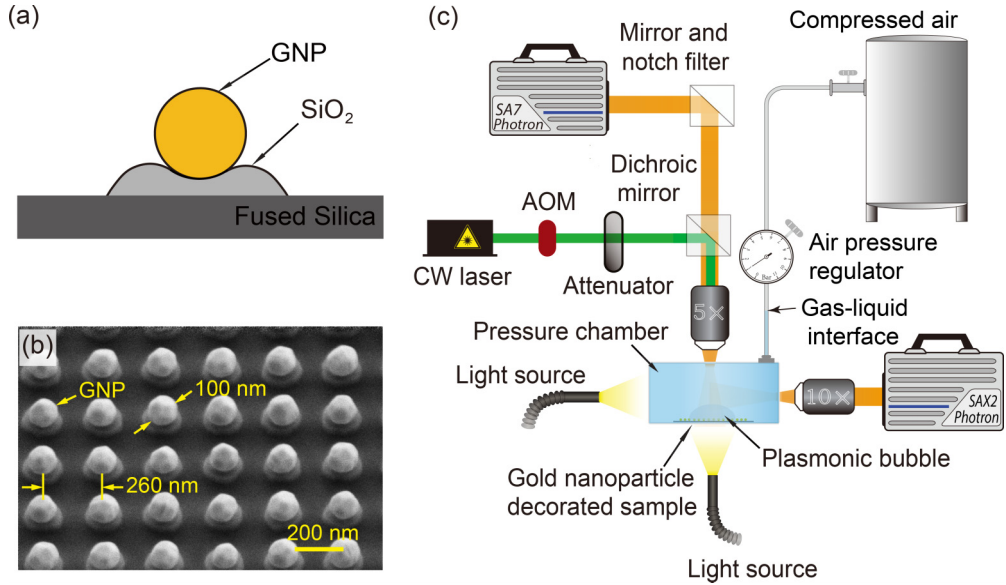

FIG. 1. (a) Schematic of a gold nanoparticle sitting on a SiO$_2$ island. (b) A scanning electron microscopy image of the gold nanoparticle decorated substrate. (c) Schematic of the optical-imaging facilities for giant initial bubble observation under different ambient pressures. A pressure chamber is used to tune the pressure from 1 to 9 bar. A narrow tube together with an elevated gas-liquid interface significantly slows down gas diffusion from the compressed air to the water in the pressure chamber. As a result, the gas concentration of the water in the pressure chamber remains almost constant throughout the experiments.

## II. EXPERIMENTAL SYSTEM

### A. Sample preparation

A gold layer of ∼45 nm was deposited on an amorphous fused-silica wafer by using an ion-beam sputtering system (home-built T'COathy machine, MESA$^+$, Twente University). The wafer was coated with a bottom antireflection coating (BARC) layer (∼186 nm) and a photoresist (PR) layer (∼200 nm). Periodic nanocolumns with diameters of ∼110 nm were patterned in the PR layer by using displacement Talbot lithography (PhableR 100C, EULITHA) [25]. Subsequently, these periodic PR nanocolumns were transferred to the underlying BARC layer, forming 110-nm BARC nanocolumns by using nitrogen plasma etching (home-built TEtske machine, NanoLab) at 10 mTorr and 25 W for 8 min. Taking these BARC nanocolumns as a mask, the Au layer was then etched by ion-beam etching (Oxford i300, Oxford Instruments, United Kingdom) with 5-sccm Ar and 50–55 mA at an inclined angle of 5°. The etching for 9 min resulted in periodic Au nanodots supported on cone-shaped fused-silica features. The remaining BARC was stripped using oxygen plasma for 10 min (TePla 300E, PVA TePla AG, Germany). The fabricated array of Au nanodots was annealed to 1100°C in 90 min. and subsequently cooled passively to room temperature. During the annealing process, these Au nanodots reformed into spherical-shaped Au nanoparticles, as shown in Figs. 1(a) and 1(b).

### B. Setup description

Figure 1(c) shows a schematic diagram of the experimental setup used for the study of initial giant bubbles under different ambient pressures $p_0$. In the setup, the gold nanoparticle decorated substrate was placed in a home-built pressure chamber. The chamber was completely filled with deionized (DI) water (Milli-Q Advantage A10 System, Germany) and connected to the compressed air via a narrow tube. Before the experiments, the DI water was exposed to air for 24 h to obtain fully air-saturated water. The gas concentration in the DI water was measured by an oxygen meter (Fibox 3 Trace, PreSens). The measured relative air concentration level was 0.99. The pressure of the chamber was tuned by an air-pressure regulator. Here we have used seven different ambient pressures of 1, 2, 3, 4, 5, 7, and 9 bar, respectively. A continuous-wave laser (Cobolt Samba) with a wavelength of 532 nm was used for irradiation of our samples. The radius $R_l$ of the laser spot was about 12.5 $\mu$m. The laser power $P_l$ projected on the sample surface was tuned via two polarization filters and measured by a photodiode power sensor (S130C, ThorLabs). Laser pulses of 10 ms were generated by a pulse–delay generator (BNC model 565).

Two high-speed cameras were used for top view and side-view imaging, respectively. The top-view camera (SA7, Photron) was used to focus the laser on the sample surface, while the side-view camera (SAX2, Photron) was used for the observation of the formation of bubbles. The top-view camera and the side-view camera were equipped with a 5× (LMPLFLN, Olympus) and a 10× (SLMPLN, Olympus) long working distance objectives, respectively. A frame rate of 540 kfps was used for side-view imaging. A home-designed algorithm was applied to segment the acquired bubble images [26–28]. With this algorithm, the bubble volume can be extracted automatically.

## III. RESULTS AND DISCUSSION

Several images of initial giant bubbles at their maximum size at the same laser power $P_l$ of 32.7 mW, but under different ambient pressures, $p_0$, are shown in Fig. 2(a). These results show that the maximum size of the giant bubble rapidly decreases with increasing $p_0$, reflecting that with increasing ambient pressure $p_0$ the expanding has to do more work against the ambient pressure. As we have previously reported, the volume of the initial phase giant bubbles is directly related to the delay time, $\tau_d$, which is defined as the time interval between switching on the laser and the nucleation of the bubble [6]. In Fig. 2(b) a semilogarithmic plot of $\tau_d$ as a function of laser power $P_l$ is shown. As already seen in Ref. [6], the delay time, $\tau_d$, decreases with increasing laser power $P_l$, but here we find that $\tau_d$ is independent of $p_0$ for a fixed $P_l$ in the range of 1 to 9 bar; see Fig. 2(c). We noticed that the measured delay time for all three laser powers under 5 bar is relatively higher than for the other pressure values. We speculate that this is a systematic error, presumably originating from the laser spot under 5 bar being slightly out of focus, leading to a slightly lower laser power density and hence increased delay time.

Before bubble nucleation, the water has to be heated up to the nucleation temperature $T_n$, which usually substantially exceeds the boiling temperature $T_{\text{boil}}$ [29–31]. The higher the laser power $P_l$, the faster the surrounding water heats up and the shorter the delay time $\tau_d$. The nucleation temperature $T_n$ can be numerically determined; for details, see Refs. [6,22]. The spatial-temporal evolution of the temperature of water, $T(r, t)$, surrounding a gold nanoparticle that is heated by a laser can be numerically calculated by solving the heat-diffusion equation,

$$\partial_t(T(r,t)) = \frac{P_l(r,t)}{\rho c_p} + \kappa \frac{1}{r^2}\partial_r(r^2 \partial_r T(r,t)), \quad (1)$$

where $\kappa$, $\rho$, and $c_p$ are thermal diffusivity, density, and heat capacity of water, $r$ is the distance to the nanoparticle, $t$ is the time, and $P_l(r, t)$ the laser power density (in W/m$^3$). For the numerical solution of the partial differential equation (1), as spatial boundary condition we took the specific configuration of the gold nanoparticle decorated sample surface used in the experiment. The heat conductivity of water and fused silica are 0.61 and 1.38 W/(mK), respectively. This simple thermal diffusion model does not include the interfacial thermal resistance term (Kapitza), which does not play a role here because our timescale exceeds the timescale of the study reported in Ref. [32] (where it is considered) by several orders of mag-nitude. The temperature field $T(r, t)$ generated by an array of nanoparticles can be considered as the linear superposition of the temperature distribution fields of the individual gold nanoparticles within a Gaussian laser beam profile,

$$T(x, y, z, t) = \sum_{i=1}^{N_{np}} [T_i(d_{i,(x,y,z)}, t)], \quad (2)$$

where $N_{np}$ is the number of gold nanoparticles under laser irradiation, $T_i$ is the temperature field produced by the $i$th nanoparticle, and $d_i$ is the distance to the center of the $i$th nanoparticle. Note that the delay time before the initial plasmonic bubble nucleation is more than 50 $\mu$s, which is much longer than the thermal relaxation time of 10 $\sim$ 100 ps for the electrons in the metal nanoparticles mentioned in Ref. [33].

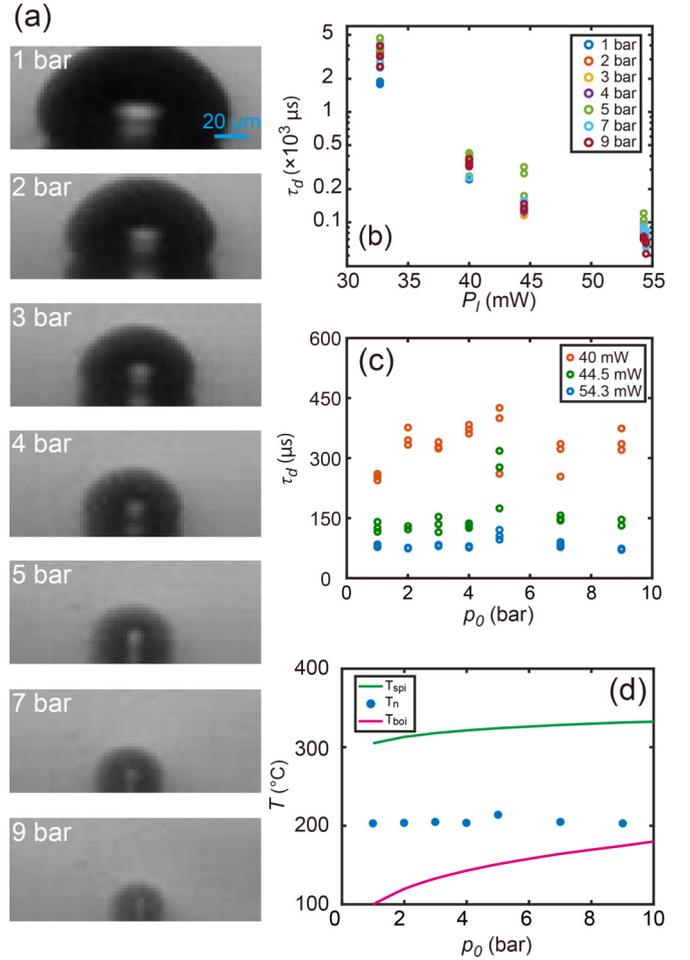

FIG. 2. (a) Examples of side-view images of initial giant bubbles at the same laser power $P_l = 32.7$ mW, but at different ambient pressures $p_0$ (see legend). The respective snapshots were taken at the maximum of the bubble expansion. (b) Delay time $\tau_d$ vs $P_l$ at different ambient pressures. (c) Delay time $\tau_d$ vs $p_0$ at different values of $P_l$ (see legend). (d) Bubble nucleation temperature $T_n$ vs the ambient pressure $p_0$. The nucleation temperature $T_n$ is obtained by fitting $\tau_d$ with the numerical model. It is found to be independent of $p_0$. We also show the spinodal temperature $T_{\text{spi}}$, i.e., the theoretical maximal attainable temperature of the liquid without vapor bubble nucleation.

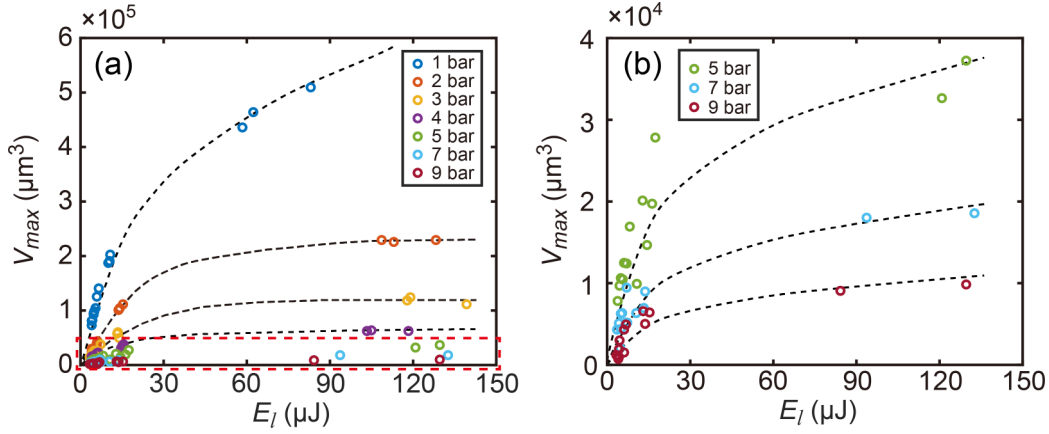

FIG. 3. (a) Maximum volume $V_{\max}$ of the bubbles as a function of the total deposited laser energy $E_l$ under different ambient pressures $p_0$ (see legend) ranging from 1 to 9 bar. (b) Zoomed-in version of the same plot curves for $p_0 = 5$, 7, and 9 bar [red dashed box in (a)]. All curves exhibit two regimes, namely a linear regime and a nonlinear regime, irrespective of the value of $p_0$. The dashed lines are drawn to guide the eye.

Therefore, the thermal relaxation effect of the GNPs close to the border of the laser beam can be neglected in this system.

As shown in Ref. [6], by numerically solving Eqs. (1) and (2), one can directly obtain the time required to reach the nucleation temperature of water at a given laser power. This approach was employed for all the experimental data using a root-mean-square minimization method. In this way, $T_n$ was obtained for different ambient pressures and is shown in Fig. 2(d). Interestingly, $T_n$ is independent of the ambient pressure and has values around 200°C.

In addition, the results shown in Figs. 2(b)–2(d) also provide insight into the dependence of $T_n$ on the amount of dissolved gas in water. Previous studies have shown that $\tau_d$ strongly depends on the gas concentration in water [6,34]. In the experiments, the absolute gas concentration is independent of the ambient pressure, $p_0$, as we do not give the water the time to be equilibrated after changing the ambient pressure. We, therefore, arrive at the conclusion that the nucleation temperature mainly depends on the absolute amount of dissolved gas in the water, which here does not depend on $p_0$.

The maximum volume $V_{\max}$ of the bubbles as a function of the total deposited energy $E_l = P_l \tau_d$ for different values of $p_0$ is shown in Fig. 3(a). Figure 3(b) shows three curves [enclosed by the red dashed box in Fig. 3(a)] for ambient pressures of 5, 7, and 9 bar, respectively. One can see that, regardless of the exact value of $p_0$, all curves exhibit a qualitatively $V_{\max}(E_l)$ dependence. When $E_l$ is smaller than 20 µJ, $V_{\max}$ linearly increases with $E_l$, which is consistent with our previous study [6,34]. However, when $E_l$ is larger than 20 µJ, $V_{\max}(E_l)$ dependence becomes nonlinear.

In the linear regime, the amount of water vapor in the bubbles is proportional to $E_l$. The proportionality factor $k$ between energy and maximum bubble volume can be used to estimate the light-vapor conversion efficiency $\eta$. The linear regime of the $V_{\max}(E_l)$ curves for different values of $p_0$ are shown in Fig. 4(a). The extracted proportionality factor $k$ as a function of $p_0$ is shown in Fig. 4(b). It can be seen that $k$ rapidly decreases from $1.9 \times 10^4\,\mu\text{m}^3/\mu\text{J}$ to $440\,\mu\text{m}^3/\mu\text{J}$ when $p_0$ is increased from 1 to 9 bar.

We now define the efficiency $\eta$ as the ratio of the energy $E_b$ used for water vaporization during vapor bubble formation to the energy $E_d$ deposited in the gold nanoparticles before nucleation of the bubble, i.e., $\eta = E_b/E_d$. Considering a gold nanoparticle coverage of $\xi = 11.6\%$, we have $E_d = \xi E_l$, where $E_l$ is the total deposited laser energy on the sample surface. The value $\eta$ can then be written as

$$\eta = \frac{E_b}{\xi E_l}. \quad (3)$$

The energy $E_b$ required to vaporize the water is composed of two components. One component is the energy needed to heat the water to vaporization temperature and the other component deals with the phase transition of the water from liquid to vapor, i.e., the latent heat $H_{\text{vap}}$. Consequently, $E_b$ for a vapor bubble is given by

$$E_b = \left(\int_{T_0}^{T_{\text{sat}}} c_p dT + H_{\text{vap}}\right) \frac{M p_{\text{sat}} V_{\max}}{R_g T_{\text{sat}}}, \quad (4)$$

where $T_0$ and $T_{\text{sat}}$ are the ambient temperature and saturation temperature of water, respectively. $M$ is the molar mass of water (18 g/mol) and $p_{\text{sat}}$ is the saturation pressure of water vapor at the moment that the bubble reaches its maximum volume. $V_{\max}$ is the maximum volume of the bubble and $R_g = 8.314\,\text{J}/(\text{mol K})$ the universal gas constant. By combining Eqs. (3) and (4), we find

$$\eta = \left(\int_{T_0}^{T_{\text{sat}}} c_p dT + H_{\text{vap}}\right) \frac{M p_{\text{sat}} V_{\max}}{R_g T_{\text{sat}} \xi E_l}. \quad (5)$$

To calculate $\eta$ from Eq. (5), we note that the saturation pressure $p_{\text{sat}}$ is close to the ambient pressure $p_0$ and can be estimated to be $p_0 - 0.04$ bar [6]. Once $p_{\text{sat}}$ is determined, $T_{\text{sat}}$ can be obtained [Fig. 4(c)] [35]. The ratio $p_{\text{sat}}/T_{\text{sat}}$ is dependent on $p_0$. The efficiency can be obtained using the prefactor $k$ for the linear regime in the $V_{\max}(E_l)$ dependence at a given $p_0$. The obtained efficiency as a function of $p_0$ is shown in Fig. 4(d). The efficiency decreases from 25 to 5% when $p_0$ is increased from 1 to 9 bar.

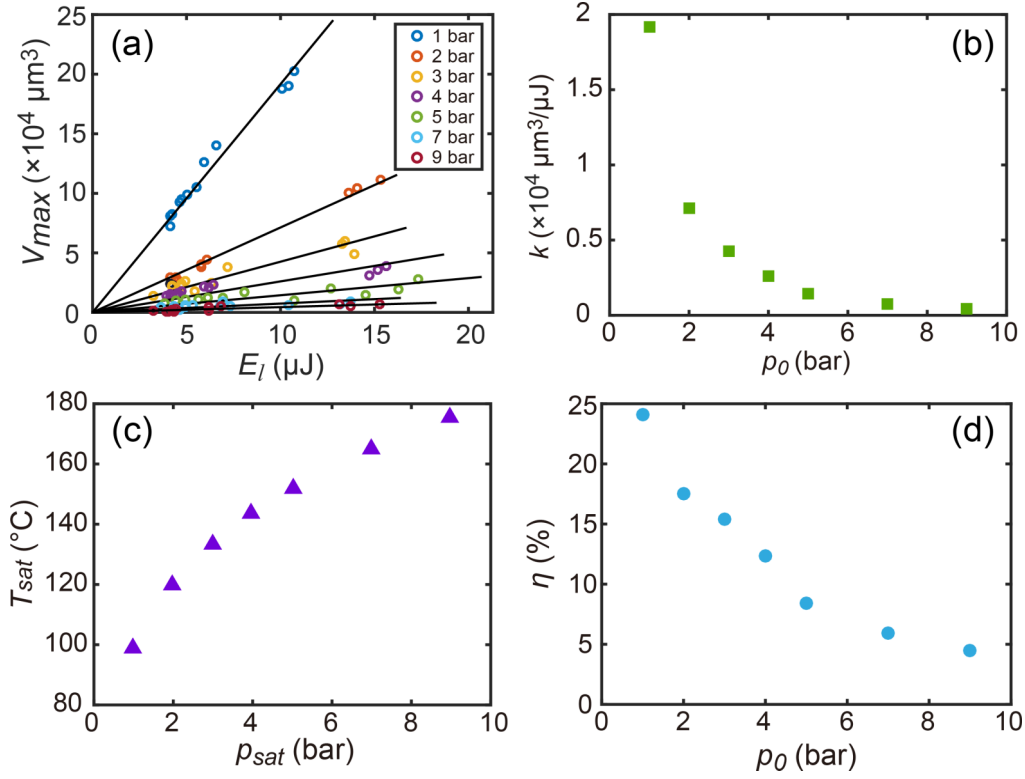

FIG. 4. (a) The maximum volume $V_{max}$ of the giant bubbles as a function of $E_l$ in the linear regime under different ambient pressures $p_0$ (see legend). (b) The prefactor $k = V_{max}/E_l$ of the linear relation $V_{max}$ vs $E_l$ as a function of $p_0$. (c) Saturation temperature $T_{sat}$ vs saturation pressure, $p_{sat}$, at the moment of maximum giant bubble volume under different ambient pressure $p_0$. (d) Experimentally obtained light-vapor conversion efficiencies $\eta$ vs the ambient pressure $p_0$.

In the nonlinear regime of the $V_{max}(E_l)$ dependences, $k$ is still defined as $V_{max}/E_l$ and obviously depends on $E_l$ and also $p_0$. Following Eq. (5), the light-vapor conversion efficiency changes accordingly. For the experimental results shown in Fig. 3, the corresponding efficiency as a function of laser power $P_l$ and ambient pressure $p_0$ is presented in Fig. 5, revealing that the efficiency decreases with increasing $p_0$ and decreasing $P_l$.

To better understand how $P_l$ and $p_0$ affect $\eta$ during the nucleation of the initial phase giant bubbles, we numerically solve Eqs. (1) and (2) for a whole range of $P_l$ and $p_0$. An example of the constructed temperature distribution field is shown in Fig. 6(a). From this figure, one can see that the temperature of the water rapidly decreases with increasing distance away from the center of the laser spot. In our model, we assume that the following two conditions are valid during the nucleation of the bubble: (1) the bubble starts to nucleate when the highest temperature of the surrounding water has reached the nucleation temperature $T_n$, (2) the volume of the bubble is determined by the amount of water that has a temperature higher than an ambient pressure-dependent threshold temperature, which is defined as vaporization temperature $T_{vap}$. In Fig. 6(b), a zoom-in plot of the temperature distribution of the red dashed box in Fig. 6(a) is shown. The key question is of course how to determine $T_{vap}$.

Given a certain water temperature distribution, the value $T_{vap}$ determines the amount of water that can be vaporized in case a bubble nucleates. A higher value of $T_{vap}$ implies a smaller volume of water and thus a smaller bubble. Therefore, the maximum size of the bubble allows us to determine $T_{vap}$. The amount of moles of vaporized water molecules $n_{max,exp}$ in a giant bubble is given by

$$n_{max,exp} = \frac{p_{sat} V_{max}}{R_g T_{sat}}. \quad (6)$$

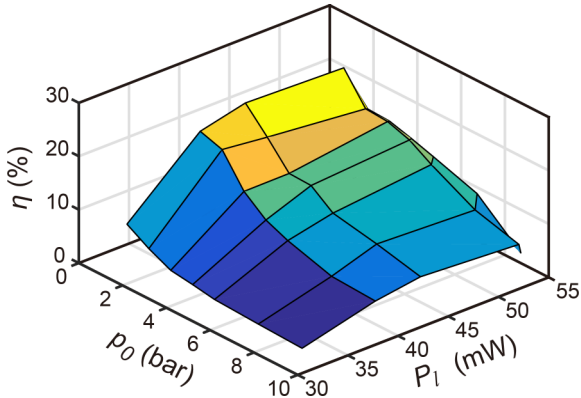

FIG. 5. The experimentally measured light-vapor conversion efficiency, $\eta$, as a function of laser power $P_l$ and ambient pressure $p_0$.

$n_{max,exp}$ as a function of $E_l$ for different pressures $p_0$ is plotted in Fig. 6(c) (circles refer to the experimental data). It clearly shows that for a given laser energy $E_l$ the amount of vaporized water decreases with increasing $p_0$. For higher values of the laser energy $E_l$, $n_{max,exp}$ levels off. The above two observations are consistent with the obtained $\eta(P_l)$ and $\eta(p_0)$ dependences.

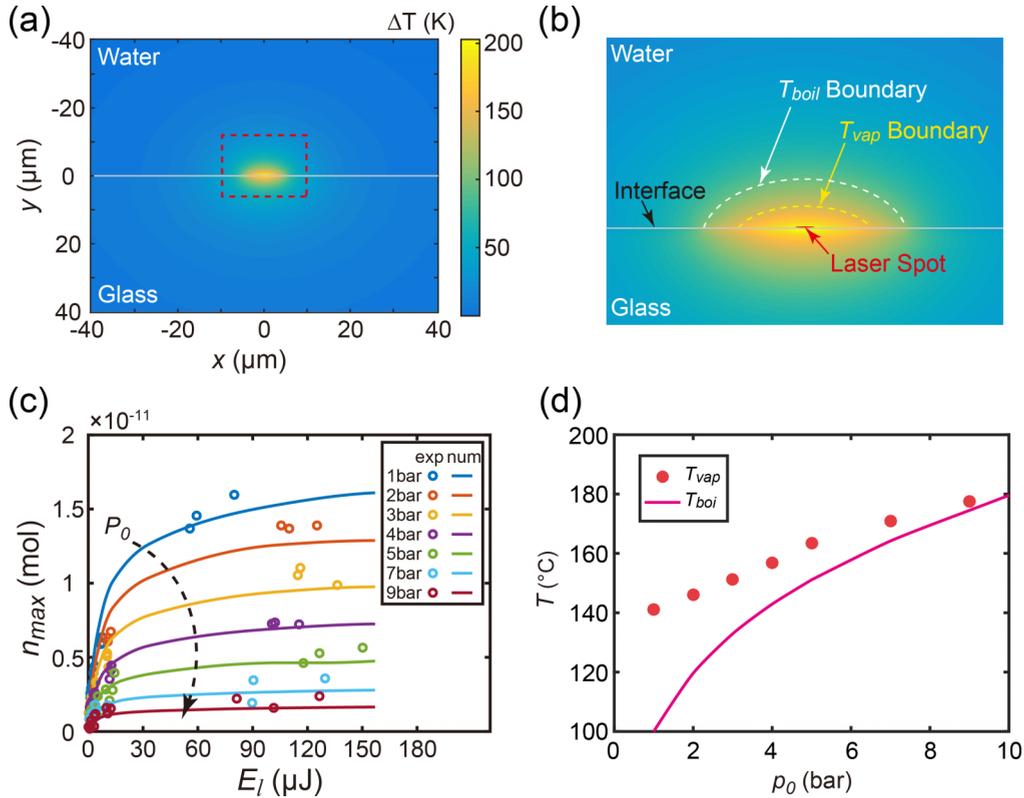

FIG. 6. (a) Water temperature field in the vicinity of a laser spot constructed by numerically solving the model discussed in the text. (b) Enlarged view of the water temperature field [dashed red box in panel (a)]. Once the water temperature at the impact point of the laser reaches $T_n$, water within the regime with boundary of $T_{vap}$ rapidly vaporizes and a giant bubble is nucleated. (c) Experimentally measured (data points) and numerically calculated (curve) amount of vaporized water in moles, $n_{max}$, as a function of the total deposited laser energy $E_l$ for various ambient pressures $p_0$ (see legend). (d) Vaporization temperature $T_{vap}$ and boiling point $T_{boil}$ vs ambient pressure $p_0$.

The thermal diffusivity $\kappa$ and specific heat capacity $c_p$ of water only very weakly depend on the ambient pressure (Table I in the Appendix). We have shown that the nucleation temperature $T_n$ is independent of the ambient pressure. Based on these dependences the water temperature distribution is independent of the ambient pressure $p_0$ for a fixed laser power $P_l$. The temperature distribution in the water depends, however, on the laser power. Upon selecting a value for $T_{vap}$ the amount of water molecules can be calculated, which implies that we can extract $n(T_{vap})$ for each laser power. The total deposited laser energy in the numerical calculation is given by $E_l = P_l \tau_d$. Subsequently, the $n(E_l)$ dependence can be obtained by gradually tuning $T_{vap}$ from room temperature to $T_n$ for each value of $P_l$. Using the $n(E_l)$ dependence, $T_{vap}(p_0)$ can be determined by minimizing $\Sigma[n(E_l) - n_{max,exp}(E_l)]^2$. The results are displayed in Fig. 6(c). The numerically determined $n_{max}(E_l)$ dependence agrees well with the experimentally obtained results.

The numerically determined $T_{vap}$ for different ambient pressures is shown in Fig. 6(d). The solid circles refer to the numerically determined $T_{vap}$, while the solid curve represents the water boiling point $T_{boil}$ as a function of $p_0$. It is clear that $T_{vap}$ is in between $T_{boil}$ and $T_n$ [around 200°C, as shown in Fig. 2(d)]. With increasing $p_0$, the vaporization temperature $T_{vap}$ gets closer to $T_n$.

We now return to the observed dependences of $\eta$ on the laser power $P_l$ and the ambient pressure $p_0$. A higher laser power $P_l$ leads to a faster increase of the water temperature and to a short delay time $\tau_d$. If $\tau_d$ is small compared to the thermal diffusion timescale $\tau_{diff} \approx R_l^2/(\pi \kappa)$, only a small amount of energy can diffuse into the nonvaporizable zone, resulting in a high efficiency. On the contrary, a lower laser power $P_l$ leads to a longer delay time $\tau_d$. Since the thermal diffusivity $\kappa$ is almost independent of the ambient pressure, an increased delay time $\tau_d$ results in an increased amount of energy diffusion into the nonvaporization zone and hence a lower efficiency. Regarding the ambient pressures $p_0$, a higher value will lead to an increased vaporization temperature $T_{vap}$. As a result, a reduced portion of heated water will be vaporized. Although the delay time of bubble nucleation remains constant for different ambient pressures, the portion of laser energy used for water vaporization decreases, resulting in a decreased light-vapor conversion efficiency.

## IV. CONCLUSIONS

We have systematically investigated the nucleation of initial giant plasmonic bubbles in water with boiling points ranging from 100 to 175°C by tuning ambient pressure from 1 to 9 bar. The experimental observations can be quantitatively understood within a theoretical framework based on the thermal diffusion equation and the thermodynamics of the phase transition. It has shown that water in the vicinity of laser-irradiated gold nanoparticles can be divided into a vaporization zone and a nonvaporization zone. The two zones are divided by vaporization temperatures, above which water will be vaporized during the giant bubble nucleation. Water in the vaporization zone vaporizes when the bubble nucleation temperature is reached. This bubble nucleation temperature only depends on the absolute amount of gas dissolved in the water, while the vaporization temperature increases with water boiling points. As a result, the light-vapor conversion efficiency decreases with increasing boiling points.

This study of the light-vapor conversion efficiency of laser-irradiated Au nanoparticles in water is also relevant for applications. For example, noble-metal nanoparticles are one of the most commonly used solar energy absorbers. Our study demonstrates that the interfacial (localized) heating can significantly increase the solar-vapor conversion efficiency [36–38].

### ACKNOWLEDGMENTS

This work is partially supported by National Natural Science Foundation of China (Grants No. 51775028 and No. 52075029) and Beijing Natural Science Foundation (Grant No. 3182022). The authors thank the Dutch Organization for Research (NWO) and the Netherlands Center for Multiscale Catalytic Energy Conversion (MCEC) for financial support. D.L. acknowledges financial support by an ERC Advanced Grant "DDD" under Project No. 740479 and by NWO-CW. Y.W. appreciates the financial support from Beijing Youth Talent Support Program, and B.Z. thanks the Chinese Scholarship Council (CSC) for financial support.

### APPENDIX: PHYSICOCHEMICAL PROPERTIES OF WATER UNDER DIFFERENT AMBIENT PRESSURES

The physicochemical properties of pure water under ambient pressures of 1 bar and 10 bar are listed in Table I. The results show that density $\rho$, thermal conductivity $\lambda$, thermal diffusivity $\kappa$, latent heat of vaporization $H_{\text{vap}}$, and specific-heat capacity $c_p$ of pure water at 10 bar are very close to that at 1 bar. Therefore, we can assume that the above 4 parameters of water basically remain constant when ambient pressure changes from 1 to 9 bar.

TABLE I. Physicochemical properties of water under ambient pressure of 1 and 10 bar [35]. All values are taken at 25°C.

| Parameters | 1 bar | 10 bar |
| --- | --- | --- |
| Density $\rho$ (kg/m$^3$) | 997.05 | 1000.3 |
| Thermal conductivity $\lambda$ [mW/(m K)] | 606.52 | 610.0 |
| Thermal diffusivity $\kappa$ ($\times 10^{-6}$ m$^2$/s) | 0.146 | 0.145 |
| Latent heat of vaporization $H_{\text{vap}}$ (kJ/kg) | 104.92 | 113.48 |
| Specific-heat capacity $c_p$ [kJ/(kg K)] | 4.1813 | 4.1973 |